*Research Article*

# Pickling Behaviour of 2205 Duplex Stainless Steel Hot-rolled Strips in Sulfuric Acid Electrolytes

**Jianguo Peng**  [1, 2], **Moucheng Li** [1*]

[1] Institute of Materials, Shanghai University, Shanghai 200072, China;
[2] Research Institute, Baoshan Iron & Steel Co., Ltd., Shanghai 200431, China

Correspondence should be addressed to Jianguo Peng; peng_jianguo@baosteel.com and Moucheng Li:mouchengli@shu.edu.cn

Pickling behaviour of the oxide layer on hot-rolled 2205 duplex stainless steel (DSS) was studied in $H_2SO_4$ solutions with electrolytic workstation. The pickling rate at 75 ºC increases slightly with the $H_2SO_4$ concentration from 100 to 250 g $L^{-1}$, but increases markedly as the concentration is higher than 250 g $L^{-1}$. The solution temperature greatly accelerates the pickling rate. $Fe^{3+}$ ion is an effective oxidant, which shows stronger inhibitive effect. The rotating disk electrode was used to simulate the moving state of steel strips in practice. The movement with the speed from 0 to 20 m $min^{-1}$ results in weak acceleration effect on the pickling process. Under dynamic conditions, the pickling rate increases noticeably with changing the pulse current density from 0 to 0.2 A $cm^{-2}$. The industrial pickling efficiency of 2205 DSS hot-rolled strips increases from 5-8 m/min to 15-18 m/min. The surface quality is improved.

## 1. Introduction

Type 2205 duplex stainless steel (DSS) has been used widely in many industries for its excellent mechanical properties and corrosion resistance [1-6]. However, the oxide scales formed on the surface of 2205 DSS during rolling and heat treatment are stable and dense due to the high chromium, molybdenum and nitrogen contents [7]. Moreover, oxide nodules may often grow on the surface of 2205 DSS for the different oxidation rates of ferritic and austenitic phases at high temperature [8-9]. Therefore, the pickling of 2205 DSS hot-rolled strips is more difficult than that of standard austenitic grades [10].

The pickling process of stainless steels is used to remove the oxide scales and chromium depleted layer in order to restore the corrosion resistance and ensure the surface finish [11-12]. The industrial pickling process includes mechanical descaling, pre-pickling, and mixed-acid pickling. A mechanical descaling is achieved through breaking, grinding, roto blasting, and sand blasting processes. Pre-pickling is performed either anodically or cathodically in acid or neutral electrolytes with certain current densities [13-15]. The general electrolytes are $Na_2SO_4$, $H_2SO_4$, $H_3PO_4$, HCl, HF and $H_2O_2$. Mixed-acid pickling is the final step to dissolve the oxide scales in $H_2SO_4$-HF [16], $H_3PO_4$-$H_2SO_4$ [17], HCl-HF [18] or $HNO_3$-HF [19].

Pre-pickling plays a critical role for the industrial production process of 2205 DSS hot-rolled strips. $Na_2SO_4$

electrolytes can conduct electricity, but they cannot dissolve the oxides and matrix [20]. $H_2SO_4$ solutions can dissolve the oxide scales and matrix [16, 21]. $H_3PO_4$ solutions have good properties of dissolving oxide scales, but their cost are expensive [22]. HCl solutions can dissolve ferrous oxides, but chloride corrosion is the big problem [18, 23-25]. HF solutions can dissolve oxide scales and $SiO_2$, but there is a serious intergranular corrosion problem [16,18].The dissolving effect of $H_2O_2$ is remarkable on the Cr-depleted layer of stainless steel, but unsatisfactory on the oxide scale [26-27]. After comparison, $H_2SO_4$ solutions are extensively used in practice for their low cost and chemical stability.

Pickling process can be simulated with static and dynamic conditions at the laboratories. There are many studies on static pickling [16-26], but very few on dynamic pickling [22]. The stainless steel strips in the industrial pickling process is moving. So the studies focused on dynamic pickling are of practical significance.

The pickling technology and mechanism are well known for the austenitic and ferritic stainless steels in the literature, but there are few systematical reports focused on the pickling for duplex stainless steel, especially for DSS 2205 hot-rolled strips. With the rapid development of economy and living standard, the demand for 2205 DSS is increasing quickly. However, the production of 2205 DSS is restricted greatly by its pickling process. It is necessary to study the pickling laws and improve the pickling efficient for 2205 DSS hot-rolled strips. In this work, the pickling behaviour of 2205 DSS hot-rolled strip in $H_2SO_4$ electrolytes under both static and dynamic conditions was investigated. The pickling mechanism was discussed on the basis of mass loss and surface analysis.

## 2. Experimental

*2.1 Materials.* A hot-rolled 2205 duplex stainless steel strip with a thickness of 4.0 mm was used as the test materials. Its chemical composition is given in table 1.

**TABLE 1.** Chemical composition of 2205 DSS specimens (wt. %)

| C | Si | Mn | S | P | Cr | Ni | Mo | N |
|---|----|----|---|---|----|----|----|---|
| 0.024 | 0.62 | 1.42 | 0.001 | 0.024 | 21.13 | 5.46 | 3.11 | 0.15 |

*2.2 Sample Preparation.* A plate of 300×300 mm was cut from the strip and annealed at 1140 ºC for 4 min. After annealing, the plate was taken out from the furnace and cooled down in air. The annealed plate was subjected to sand blasting to remove its main oxides on the surface and then was cut into test specimen of 5×5 mm. The specimens were covered with epoxy resin leaving a working area of 5×5 mm and cleaned with alcohol before pickling.

*2.3 Pickling Measurement.* Pickling tests were carried out on the workstation as shown in Figure 1. A specimen fixed on the rotating disk with a diameter of 15 cm was used as working electrode, a platinum sheet as counter electrode, and $Hg/Hg_2SO_4$ as reference electrode. The analytically pure $H_2SO_4$, ferrous sulfate, iron sulfate and de-ionized water were used to prepare the electrolytes. The pickling processes were performed with a Princeton potentiostat (PAR 273A).

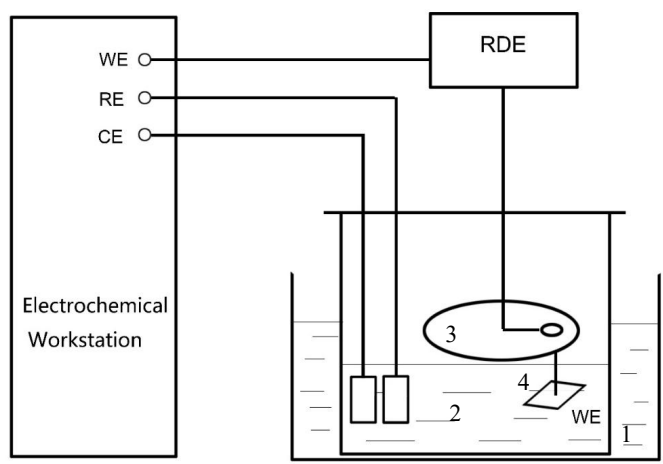

**FIGURE 1:** Electrolytic workstation used for dynamic pickling tests.
1-thermostat water bath, 2-electrolyte, 3- rotating disk, 4-specimen

*2.4 Electrochemical measurement.* The corrosion potential and electrochemical impedance spectroscopy (EIS) of the test specimens were measured during the pickling process with a Princeton potentiostat (VMP3). Each specimen was fixed on the cell sidewall with a 1 cm$^2$ round hole. The test specimen was used as the working electrode; the platinum sheet was used as the counter electrode; and the $Hg/Hg_2SO_4$ was used as the reference electrode (MSE).

All test samples were immersed in the electrolytes for 2 min. During electrolysis, the electrolytic time was 81 seconds with three periods. For each period, the specimen was polarized anodically for 18 seconds and cathodically for 9 seconds at various current densities. The pickling rate was calculated from the mass loss of specimen with a precision of 0.01 mg. The specimens were characterized with scanning electron microscopy (FEI, qunta 600) and Roman spectroscopy (Renishaw, INVIA).

## 3. Results

*3.1 Characterization of oxide scales.*

In general, the oxide scale of hot-rolled 2205 DSS strip after annealing can be divided into an inner layer and an outer layer[7]. Chromium is rich in the inner layer while iron is rich in the outer layer. The outer layer is consisted of $Fe_2O_3$, $Fe_3O_4$ and $FeCr_2O_4$ and the inner layer is composed mainly of $FeCr_2O_4$ [28]. The outer layer may be partially removed by mechanical descaling. The $FeCr_2O_4$ is the oxide with a spinel structure, which is insoluble in most industrial acids[7,29].

As shown in the Figure 2, a few oxide scales remain on the surface of test specimen after shot blasting. Roman spectrum in Figure 3 indicates that the residual oxides on the surface of 2205 DSS hot-rolled strip are composed of $Fe_2O_3$, $Fe_3O_4$ and $FeCr_2O_4$. Among them, the content of $FeCr_2O_4$ is the highest as seen from its peak intensity. This implies that the pickling of 2205 DSS hot-rolled strip is difficult.

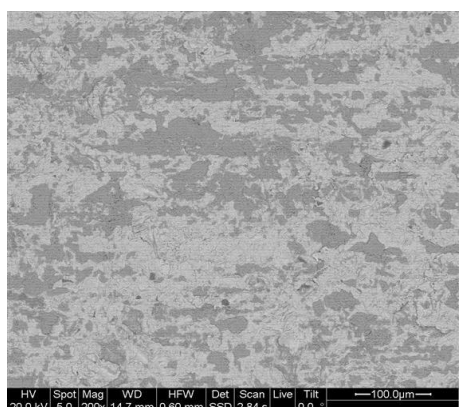 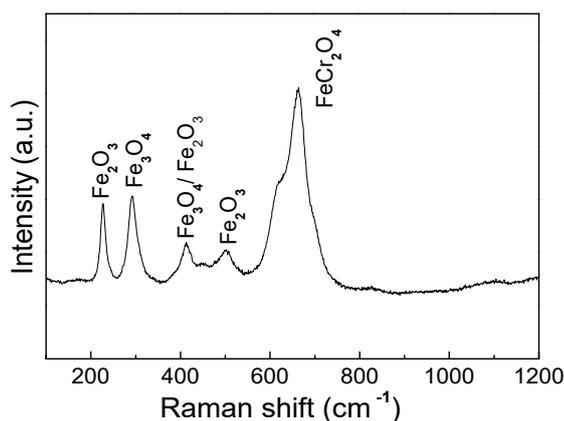

**FIGURE 2:** Surface image for the test specimen     **FIGURE 3:** Roman spectrum for the test specimens

*3.2 Chemical pickling properties under static condition.*

The static chemical pickling of stainless steel in $H_2SO_4$ solutions is mainly related with acid concentration, pickling temperature and concentration of metal ions. For most steels, higher picking temperature, higher concentration of $H_2SO_4$ solutions and lower concentration of metal ions can improve its pickling efficiency in initial pickling stage for its strong dissolving capacity. After pickling for a certain time, the strong dissolving capacity may deteriorate pickling result for the corrosion of matrix.

The effect of $H_2SO_4$ concentration on static pickling rate and typical surface images of specimens at 75 ºC are shown in Figures 4 and 5. The pickling efficiency has a positive correlation with $H_2SO_4$ concentration. It can be seen that the pickling rate has a transition at about 250 g L$^{-1}$. When the concentration increases from 100 g L$^{-1}$ to 250 g L$^{-1}$, the pickling rate is accelerated slightly and many oxide scales remain on the surface. When the concentration increases from 250 g L$^{-1}$ to 450 g L$^{-1}$, the pickling rate is accelerated markedly to remove most of the oxide scales.

Figure 6 gives the static pickling rates of specimens in 300 g L$^{-1}$ $H_2SO_4$ solutions at different temperatures. The pickling efficiency has a positive correlation with the solution temperature. With increasing temperature from 50 ºC to 90 ºC, the pickling rate increases gradually from about 1.74 to 12.8 g m$^{-2}$ min$^{-1}$ and the retained oxide scales reduced noticeably, as shown in Figure 7.

The metal ions in the $H_2SO_4$ solutions include iron, chromium, molybdenum, nickel, manganese, titanium and niobium ions. The conventional metal elements in industrial $H_2SO_4$ electrolytes are shown in table 2. Among them, iron is the main element. So the iron ions play the main impact on the pickling process.

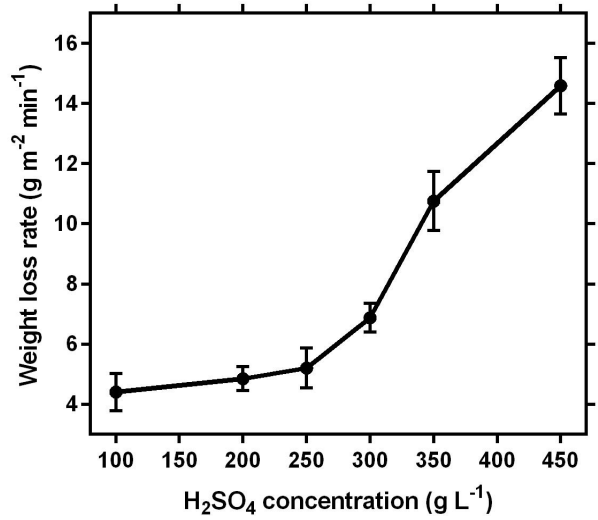

**FIGURE 4:** Variation of static pickling rate in $H_2SO_4$ at 75 °C with different concentration

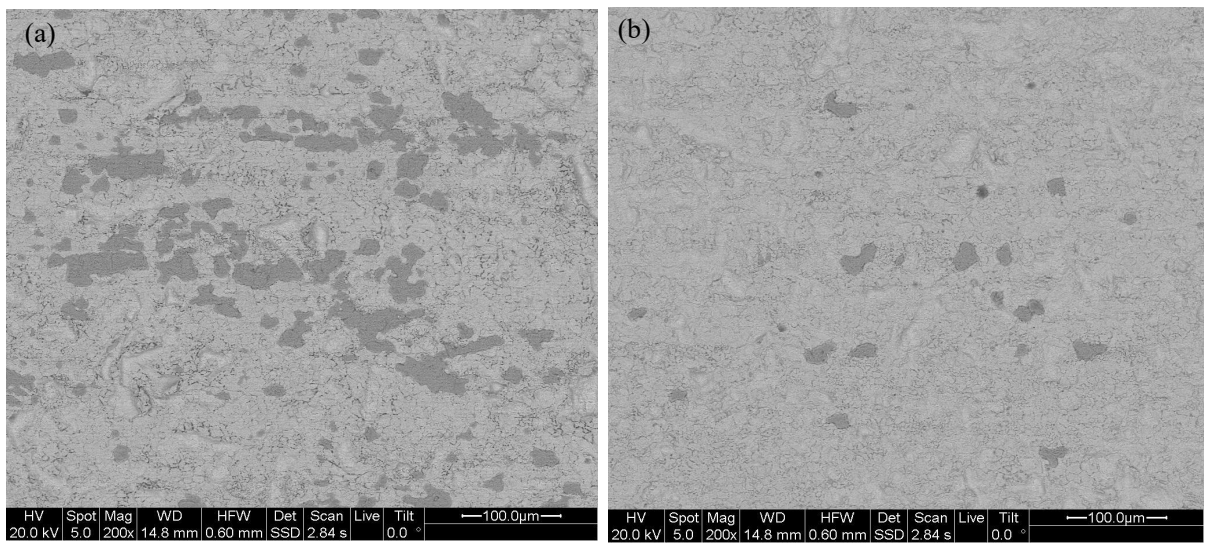

**FIGURE 5:** SEM surface images after static picking in $H_2SO_4$ solutions at 75 °C with different concentrations (a) 300 g $L^{-1}$ and (b) 450 g $L^{-1}$.

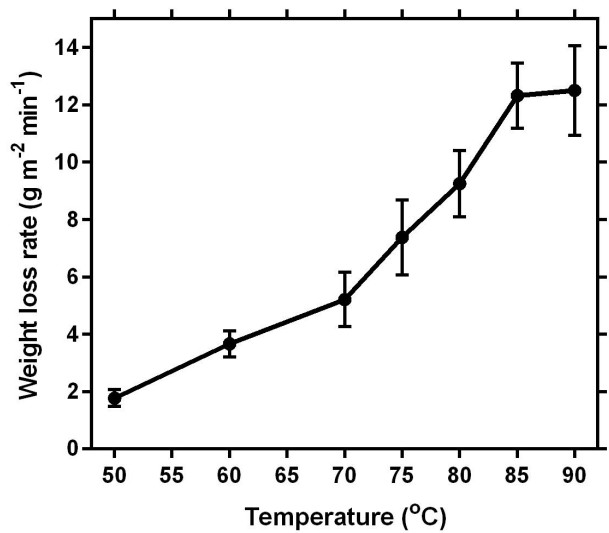

**FIGURE 6:** The static pickling rate of specimen in 300 g $L^{-1}$ $H_2SO_4$ solutions at different temperatures

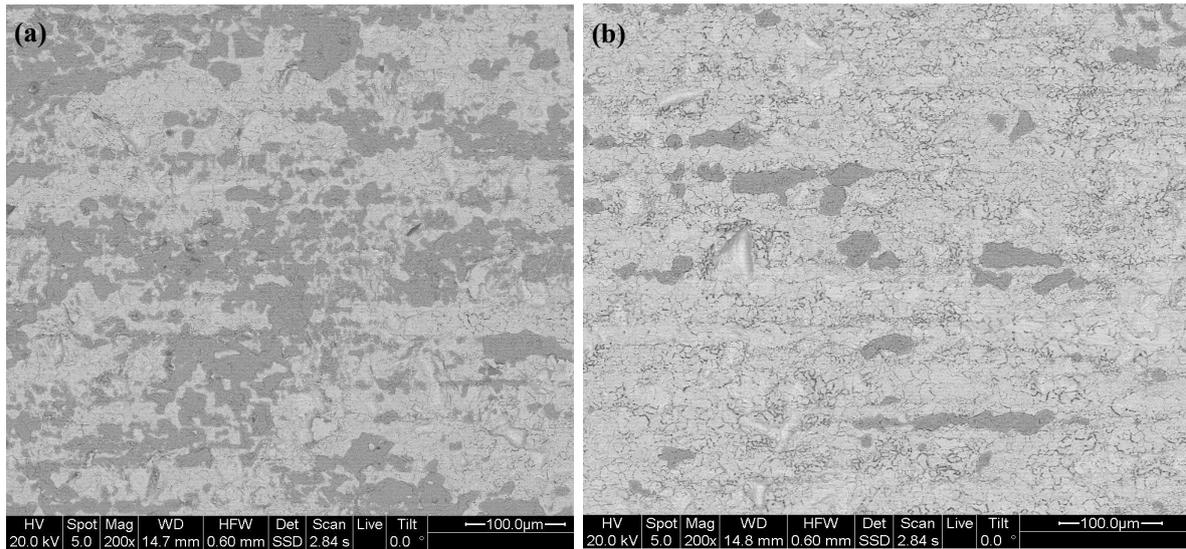

**FIGURE 7.** SEM surface images after static picking in 300 g L$^{-1}$ H$_2$SO$_4$ solutions at temperatures 60 ºC (a) and 90 ºC (b).

**TABLE 2**. Concentration of metal ions in industrial H$_2$SO$_4$ solutions

| Ions | Fe | Cr | Mn | Ni | Cu | Nb | Ti | Mo |
|---|---|---|---|---|---|---|---|---|
| Concentration (g/L) | 20.6 | 2.80 | 0.06 | 0.55 | <0.01 | <0.01 | <0.01 | 0.04 |

The effect of iron ions on the static pickling of specimen was studied in 300 g L$^{-1}$ H$_2$SO$_4$ solutions at 75 ºC and the results are given in Figure 8. The pickling efficiency has a negative correlation with concentration of iron ions. With increasing concentration of ferric ions from 0.5 g L$^{-1}$ to 60 g L$^{-1}$, the pickling rate declines sharply at first to less than 1 g m$^{-2}$ min$^{-1}$ and then decreases insignificantly. With increasing concentration of ferrous ions, the pickling rate decreases slowly, but shows higher values in comparison with ferric ions. The ferric ions are an effective oxidant which can enhance the corrosion potential of test specimens [16], then reduce their dissolution rate remarkably. In order to improve the pickling rate of 2205 DSS hot-rolled strips, new sulfuric acid solutions without iron ions are necessary.

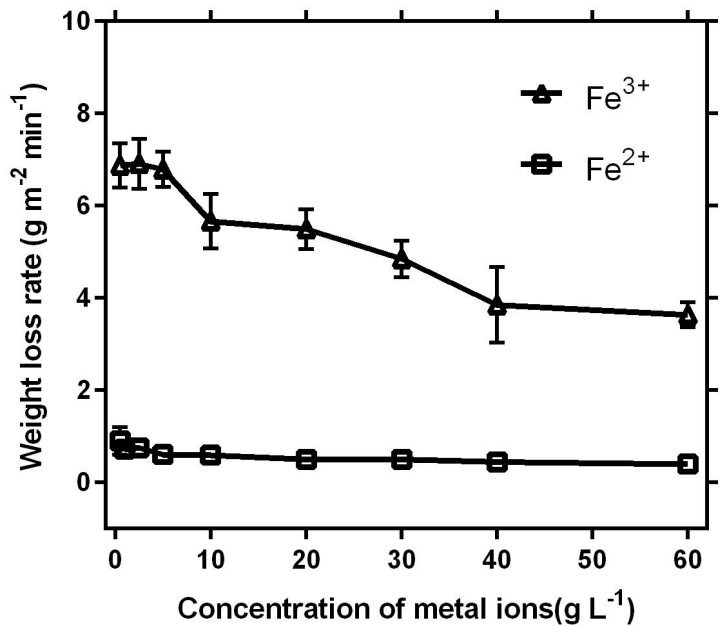

**FIGURE 8**. The static pickling rate in 300 g L$^{-1}$ H$_2$SO$_4$ solutions at 75 ºC with different iron ions

*3.3 Chemical pickling properties under dynamic conditions.*

In order to see the difference between dynamic and static pickling of hot-rolled 2205 DSS strips, the rotating disk was applied to create the movement between specimens and acid solutions. The variation of pickling rate of specimen in 300 g L$^{-1}$ H$_2$SO$_4$ solutions at 75 ºC with rotary speed is given in Figure 9. The pickling efficiency has a positive correlation with moving speed of specimen. The pickling rate is higher in the dynamic cases than in the static case. However, with the increase of rotary speed from 5 to 20 m min$^{-1}$, the pickling rate only enlarges slightly from about 7.5 to 7.8 g m$^{-2}$ min$^{-1}$. It is evident that the acceleration effect of dynamic speed is very weak.

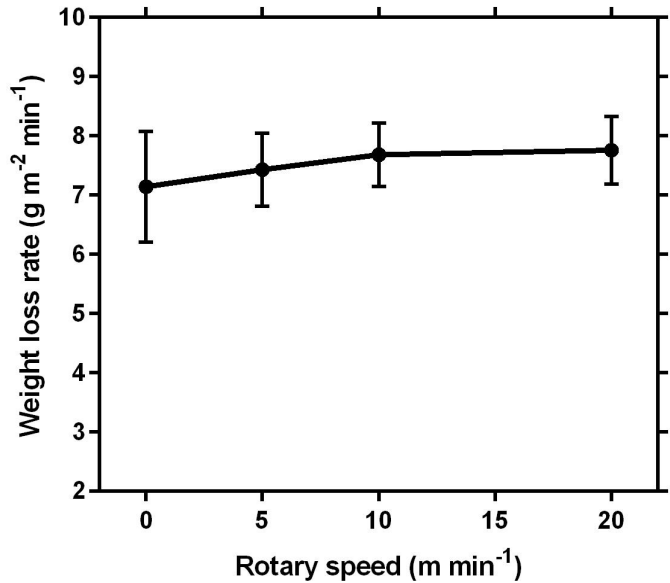

**FIGURE 9:** Dynamic pickling rate in 300 g L$^{-1}$ H$_2$SO$_4$ solutions at 75 ºC under different rotary speed conditions

*3.4 Electrolytic pickling properties under dynamic condition.*

The applied polarization plays a role in pickling process. Figure 10 gives the variation of dynamic pickling rate at 10 m min$^{-1}$ for specimens in 300 g L$^{-1}$ H$_2$SO$_4$ solutions at 75 ºC with pulse current density. The pickling rate becomes higher with increasing the pulse current density from 0 to 0.2 A cm$^{-2}$. The rate is about 2.1 times larger at 0.2 A cm$^{-2}$ than at 0 A cm$^{-2}$ (i.e., the chemical pickling). The electrolytic pickling efficiency has a positive correlation with the pulse current density. After dynamic electrolytic pickling, there is almost no oxide scales on the specimen surfaces compared with static electrolytic pickling. With the increase of current density from 0.04 A cm$^{-2}$ to 0.2 A cm$^{-2}$, the Cr-depleted layer may dissolve obviously and the surfaces become smooth, as shown in Figure11.

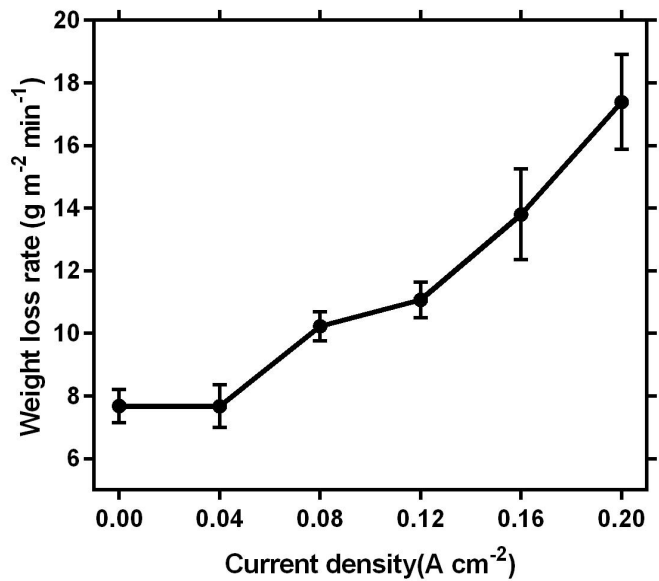

**FIGURE 10:** Variation of dynamic pickling rate with current density in 300 g L$^{-1}$ H$_2$SO$_4$ solutions at 75 ºC

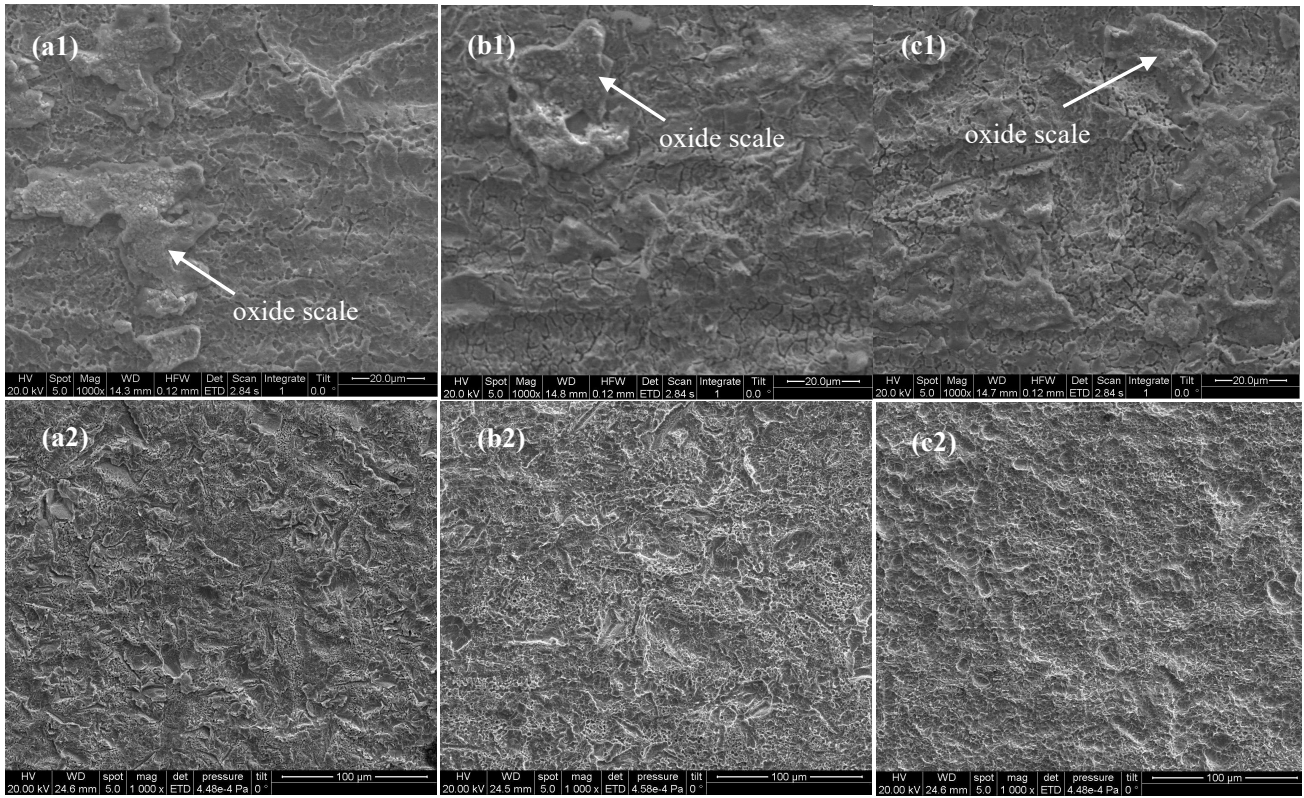

**FIGURE 11:** SEM surface images after static (a1,b1,c1) and dynamic (a2,b2,c2) picking in 300 g L$^{-1}$ H$_2$SO$_4$ solutions at 75 ºC with current density 0.04 A cm$^{-2}$ (a1,a2), 0.12 A cm$^{-2}$ (b1,b2) and 0.2 A cm$^{-2}$ (c1,c2).

## 4. Discussion

The residual oxides on the surface of test specimen are composed of Fe$_2$O$_3$, Fe$_3$O$_4$ and FeCr$_2$O$_4$ as shown in Figure 3. While pickling of 2205 DSS hot-rolled strips in H$_2$SO$_4$ electrolytes, it involves two processes, chemical pickling process and electrolytic pickling process.

When chemical pickling is carried on for 2205 DSS, FeO, Fe$_2$O$_3$, Fe$_3$O$_4$ and alloying elements can be dissolved in H$_2$SO$_4$ solutions. The chemical pickling process involves the following chemical reactions [30].

$$\text{FeO} + \text{H}_2\text{SO}_4 \rightarrow \text{FeSO}_4 + \text{H}_2\text{O} \tag{1}$$

$$\text{Fe}_3\text{O}_4 + 4\text{H}_2\text{SO}_4 \rightarrow \text{FeSO}_4 + \text{Fe}_2(\text{SO}_4)_3 + 4\text{H}_2\text{O} \tag{2}$$

$$\text{Fe}_2\text{O}_3 + 3\text{H}_2\text{SO}_4 \rightarrow \text{Fe}_2(\text{SO}_4)_3 + 3\text{H}_2\text{O} \tag{3}$$

$$\text{Fe} + \text{H}_2\text{SO}_4 \rightarrow \text{FeSO}_4 + \text{H}_2 \uparrow \tag{4}$$

$$\text{Ni} + \text{H}_2\text{SO}_4 \rightarrow \text{NiSO}_4 + \text{H}_2 \uparrow \tag{5}$$

$$\text{Cr} + \text{H}_2\text{SO}_4 \rightarrow \text{CrSO}_4 + \text{H}_2 \uparrow \tag{6}$$

The oxide scales on the test specimen are mainly comprised of FeCr$_2$O$_4$ as shown in Figure 3. The FeCr$_2$O$_4$ is insoluble in sulfuric acid solution [7]. The matrix of 2205 DSS is dissolved slowly in sulfuric acid solution for its excellent corrosion resistance. The oxide scales on the test specimen is dissolved very slowly in sulfuric acid solutions by mean of peeling mechanically. So the chemical pickling of 2205 DSS hot-rolled strip is very slowly. For most stainless steels, higher pickling temperature and sulfuric acid concentration do not increase their pickling efficiency after a certain level [30]. But for 2205 DSS, the oxide scales are still exited after chemical pickling in 450 g L$^{-1}$ H$_2$SO$_4$ at 90 C.

At the beginning of static chemical pickling process, small gas bubbles nucleate and grow on the specimen surface. During the pickling process, more and more gas bubbles will form and adsorb on the specimen surface, which must reduce the contact areas of effective pickling. As for the dynamic pickling of specimens on the rotating disk, the movement may decrease the adsorption and growth of gas bubbles on the specimen surface to a certain extent [31]. So the chemical pickling rate of the annealed 2205 DSS specimens under dynamic condition is higher than under static condition. Nevertheless, the acceleration effect of dynamic speed is very weak, with about 5~10% increase in the pickling rates as observed in Figure 8.

When sulfuric acid is used as electrolyte, it can conduct electricity. The test specimen serves as anode and cathode with the alternative change. When the test specimen serves as the relative anode, the insoluble $FeCr_2O_4$ is oxidized into soluble $HCrO_4^-$ and $Cr_2O_7^{2-}$. The anodic electrolytic reactions are as followed [10,30,32].

$$2H_2O \rightarrow O_2\uparrow + 4H^+ + 4e \quad (7)$$

$$FeCr_2O_4 + 4H_2O \rightarrow FeO + HCrO_4^- + 7H^+ + 6e \quad (8)$$

$$FeCr_2O_4 + 4H_2O \rightarrow FeO + Cr_2O_7^{2-} + 8H^+ + 6e \quad (9)$$

$$2Fe_2O_3 \rightarrow 4Fe^{3+} + 3O_2\uparrow + 12e \quad (10)$$

When the test specimen serves as the relative cathode, $Cr(OH)_3$ and $Fe(OH)_3$ are deposited on the surface of test specimen. $H_2$ is precipitated between the oxide layers, which is beneficial to the separation of the deposit sediment from the matrix. The cathodic electrolytic reactions are as followed [10,30,32].

$$2H_2O + 2e \rightarrow 2OH^- + H_2\uparrow \quad (11)$$

$$Fe^{3+} + 3OH^- \rightarrow Fe(OH)_3\downarrow \quad (12)$$

$$CrO_4^{2-} + 4H_2O + 6e \rightarrow Cr(OH)_3\downarrow \quad (13)$$

$$Cr_2O_7^{2-} + 4H_2O + 6e \rightarrow 2Cr(OH)_3\downarrow + 8OH^- \quad (14)$$

Electrolytic pickling is faster than normal chemical pickling for most stainless steels [20]. The electrolytic picking process in $H_2SO_4$ electrolytes is the combined effect of electrolysis and chemical pickling. When the electrolytic pickling is performed under static conditions, a maximum of about 20% of the current goes to dissolution reactions whereas about 80% of the current is consumed in oxygen gas production [10]. For the dense and compact oxide scales of the test specimens, spallation or peeling of the oxide scales induced by gas evolution does not play a decisive role. As shown in Figure 11, the oxide scales are remained on the test specimens after static electrolytic pickling. When the electrolytic pickling is performed under dynamic conditions, the rotary disk develops a rational way for producing active edges efficiently, where catalytic activity of the test specimens is improved and its pickling rate is accelerated [33-34]. The dynamic electrolytic pickling rate at 0.2 A cm$^{-2}$ is about 2.1 times higher than the chemical pickling, which may give rise to a smooth specimen surface without oxide scales.

The $Fe^{3+}$ ions are an effective oxidant which enhanced the corrosion potential, as shown in Figure 12. When none Fe ions were added in the $H_2SO_4$ solutions, the corrosion potential quickly declined at the beginning of the immersion time, after about 50 s, it remained constant (approximately -0.75 $V_{MSE}$) subsequently as the immersion time increased. When 10 g L$^{-1}$ $Fe^{2+}$ was added in the $H_2SO_4$ solutions, the corrosion potential quickly declined at the beginning of the immersion time, after about 100s, it remained unchanged. When 10 g L$^{-1}$ $Fe^{3+}$ was added in the $H_2SO_4$ solutions, the corrosion potential increased slowly from 0.062 $V_{MSE}$ to 0.137 $V_{MSE}$ as the immersion time increased. The EIS Nyquist plot recorded on the test specimens immersed in 300 g L$^{-1}$ $H_2SO_4$ solutions containing none ions, 10 g L$^{-1}$ $Fe^{2+}$, 10 g L$^{-1}$ $Fe^{3+}$ under the $E_{corr}$ conditions (i.e., the free corrosion states) is presented in Figure 13. The semicircle size enlarged slightly with the addition of 10 g L$^{-1}$ $Fe^{2+}$, but increased noticeably with the addition of the 10 g L$^{-1}$ $Fe^{3+}$ for its high impedance.

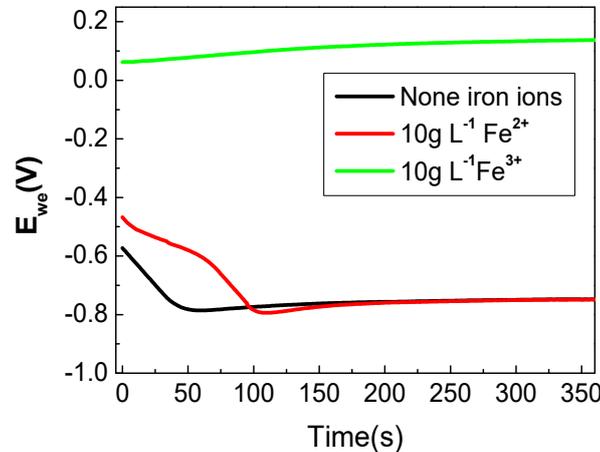

**FIGURE 12.** Corrosion potential in 300 g L$^{-1}$ $H_2SO_4$ solutions at 75 ºC containing different Fe ions

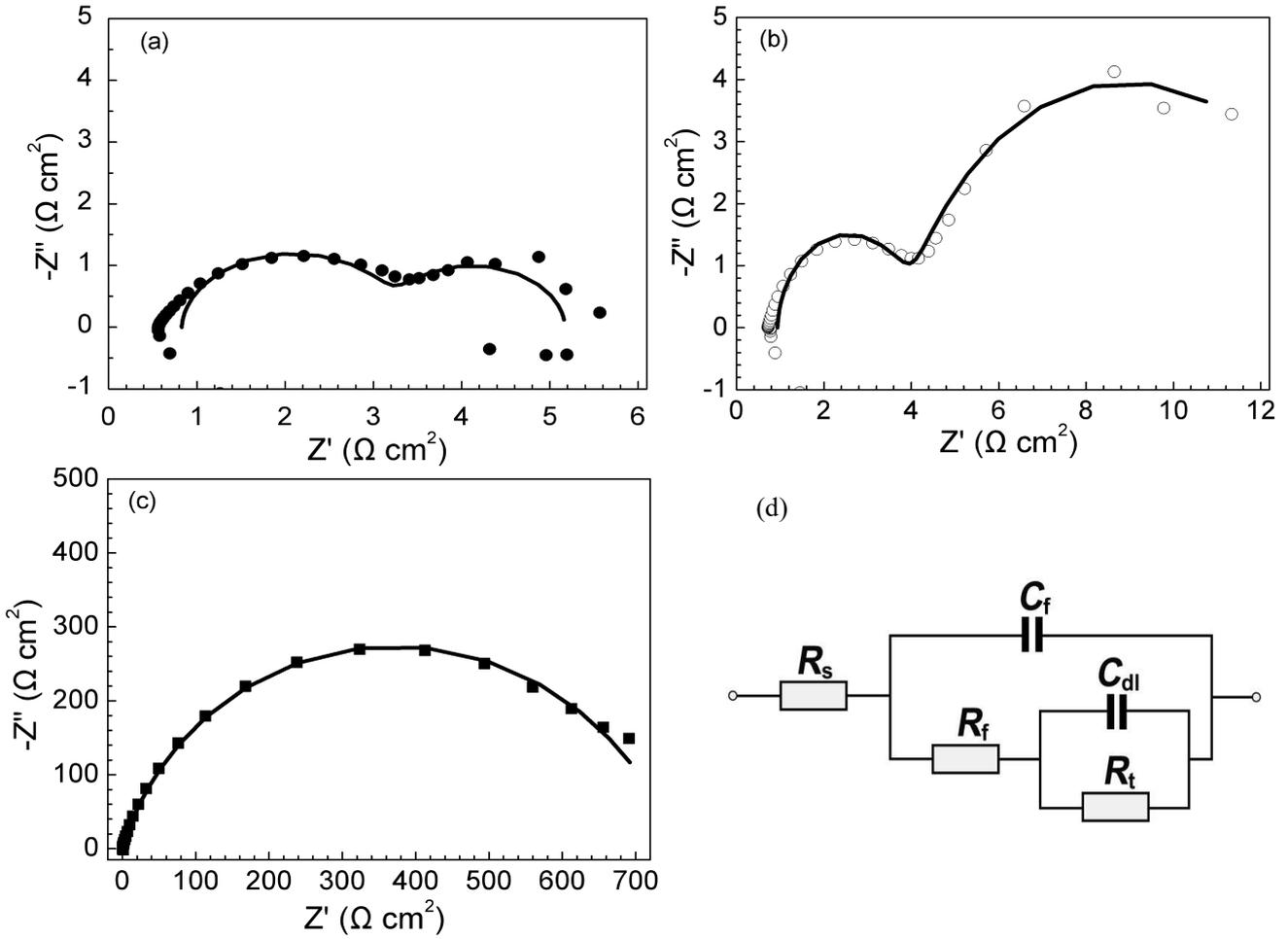

**FIGURE 13:** Nyquist plots (a,b,c) in 300 g L$^{-1}$ H$_2$SO$_4$ solutions at 75 °C containing none ions (a), 10 g L$^{-1}$ Fe$^{2+}$ (b), 10 g L$^{-1}$ Fe$^{3+}$ (c), and the corresponding circuit of the impedance spectrum (d)

In order to show the dissolution process for the specimens in 300 g L$^{-1}$ H$_2$SO$_4$ solutions at 75 °C containing different Fe ions, the equivalent circuit model is proposed in Figure 13 (d) according to the EIS features in Figure 13. $R_s$ is the electrolyte resistance. $R_f$ is the resistance of oxide layer remained on the specimens. $R_t$ is the charge transfer resistance. $C_f$ and $C_{dl}$ can be replaced with constant phase element (CPE) [35]. The impedance of CPE is written ins equation (15), where $Y_0$ is the admittance magnitude of CPE, α is the exponential term. Table 3 gives the fitted results of EIS spectra. The calculated spectra are shown as a solid curve in Figure 13, which fit the experimental data very well. It can be concluded that the model provided a reliable description for the corrosion systems.

$$Z_{CPE} = \frac{1}{Y_0(j\omega)^\alpha} \qquad (15)$$

**TABLE 3:** Fitted results for EIS spectra in 300 g L$^{-1}$ H$_2$SO$_4$ solutions at 75 °C containing different ions

| ions | $R_s$ Ω cm$^2$ | $Y_{0-f}$ S$^\alpha$ Ω$^{-1}$cm$^{-2}$ | $\alpha_f$ | $R_f$ Ω cm$^2$ | $Y_{0-dl}$ S$^\alpha$ Ω$^{-1}$cm$^{-2}$ | $\alpha_{dl}$ | $R_t$ Ω cm$^2$ |
|---|---|---|---|---|---|---|---|
| None ions | 0.82 | 0.00218 | 0.87 | 2.80 | 0.0658 | 0.92 | 1.6 |
| 10g L$^{-1}$ Fe$^{2+}$ | 0.94 | 0.00185 | 0.96 | 3.14 | 0.093 | 0.84 | 10 |
| 10g L$^{-1}$ Fe$^{3+}$ | 1.04 | 0.000133 | 0.95 | 54.99 | 0.000281 | 0.68 | 70.6 |

## 5. Application and performance

According to the above results, the pickling process of hot-rolled 2205 DSS is optimized through the high H$_2$SO$_4$ concentration, high solution temperature and proper electrolysis current density in the industrial pre-pickling of 2205 DSS hot-rolled strips. The pickling efficiency is improved remarkably, which increases the production rate from 5 8 m min$^{-1}$ to 15 18 m min$^{-1}$. Besides, the surface finish after pickling is notably enhanced, as shown in Figure 14.

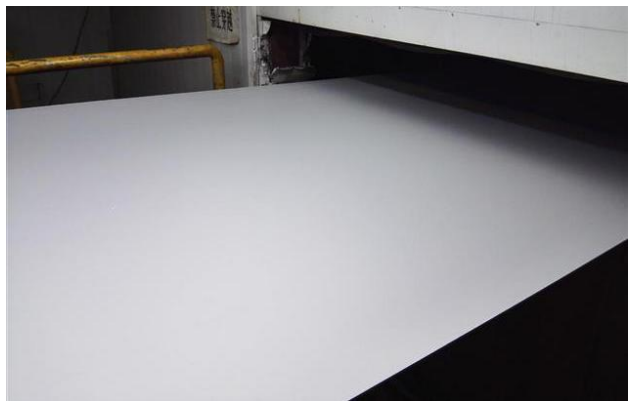

**FIGURE 14:** Optical surface of 2205 DSS hot-rolled strip after industrial pickling with optimized pickling process

## 6. Conclusions

The pickling behaviour of hot-rolled 2205 DSS with annealing and mechanical descaling treatments in $H_2SO_4$ solutions under both static and dynamic conditions has led to the following conclusions:

(1) In static chemical pickling process, the pickling rate may be accelerated noticeably by increasing the solution temperature and $H_2SO_4$ concentration, but be decelerated greatly by $Fe^{3+}$ ions.

(2) The chemical pickling process can be enhanced weakly by the moving speed of specimen from 0 to 20 m min$^{-1}$ because the movement decreases the adsorption and growth of gas bubbles on the specimen surface to a certain extent.

(3) Under dynamic conditions, the electrolytic pickling rate increases markedly with changing the pulse current density from 0.04 to 0.2 A cm$^{-2}$. The electrolytic pickling rate at 0.2 A cm$^{-2}$ is about 2.1 times larger than the chemical pickling rate in 300 g L$^{-1}$ $H_2SO_4$ at 75 ºC, resulting in the smooth and clean specimen surfaces.

## Data Availability

The tables and figures data used to support the findings of this study are available from the corresponding authors on reasonable request.

## Conflict of Interests

The authors declare that there is no conflict of interests regarding the publication of this paper.

## Acknowledgments

Financial support from National Natural Science Foundation of China (Grant Nos. U1660205 and U1960103) is gratefully acknowledged.